\documentclass[aps,showpacs,preprint]{revtex4}
\usepackage{graphicx}

\def\ket{\rangle}
\def\<{\langle}
\def\>{\rangle}
\begin{document}
\title{Modularization of the multi-qubit controlled phase gate and its NMR
implementation\footnote{Corresponding authors: Jingfu Zhang,
zhang-jf@mail.tsinghua.edu.cn  and G. L. Long,
gllong@mail.tsinghua.edu.cn}}
\author{ Jingfu Zhang$^{1,2}$, Wenzhang Liu$^{1,2}$,
Zhiwei Deng$^{3}$, Zhiheng Lu,$^{4}$
     and Gui Lu Long$^{1,2, 5}$}
\address{ $^{1}$Key Laboratory For Quantum Information and Measurements,  and
Department of Physics, Tsinghua University, Beijing,
100084, P R China\\
 $^{2}$Center For Quantum Information, Tsinghua University, Beijing 100084, P R China\\
$^{3}$Testing and Analytical Center, Beijing Normal University,
 Beijing, 100875, P R China\\
$^{4}$Department of Physics, \small{}Beijing Normal University,
Beijing, 100875, P R China \\
$^5$Center of Atomic and Molecular Nanosciences, Tsinghua
University, Beijing 100084, P R China}
\date{\today}
\begin{abstract}
Quantum circuit network is a set of circuits that implements a
certain computation task. Being at the center of the quantum
circuit network, the multi-qubit controlled phase shift is one of
the most important quantum gates. In this paper, we apply the
method of modular structuring in classical computer architecture
to quantum computer and give a recursive realization of the
multi-qubit phase gate.  This realization of the controlled phase
shift gate is convenient in realizing certain quantum algorithms.
We have experimentally implemented this modularized multi-qubit
controlled phase gate in a three qubit nuclear magnetic resonance
(NMR) quantum system. The network is demonstrated experimentally
using line selective pulses in NMR technique. The procedure has
the advantage of being simple and easy to implement.
\end{abstract}
 \maketitle

\section{Introduction}
\label{s1}

   Quantum computer is born with the combination of quantum mechanics and computer
science. Quantum properties have enabled quantum computer to
outperform a classical computer for factorizing a large number
\cite{shor}, or searching an unordered database  \cite{Grover97}.
In Ref. \cite{Deutsch85}, Deutsch  proposed quantum gates and
quantum network, and described a blue print for quantum computer.
It has been shown that it is possible to construct arbitrary
$n$-qubit quantum gate by using only a finite set of one-qubit
gates and two-qubit gates \cite{Deutsch95, Bremner}. These basic
quantum gates are universal for quantum computation\cite{Nielsen}.
Barenco et al, and  Cleve have developed methods for  designing
networks in multiple- qubit system \cite{Barenco, Cleve}.

 Quantum network can be viewed as the quantum
analog of the classical computing. It provides with us a
convenient method to  build a quantum computer similar to building
a classical computer. A classical compute can be implemented as a
fixed classical gate array, with an input programm and data. A
universal gate array can be programmed to perform any possible
function on the input data. However for a quantum computer,
Nielsen and Chuang demonstrated a rather different property
\cite{Nielsen97}. They pointed out it is not possible to build a
fixed, general purpose quantum computer which can be programmed to
perform any arbitrary quantum computation task. The implementation
of a universal quantum gate array on a quantum computer can only
be realized in a probabilistic fashion, not in a deterministic
fashion. By extending the one-to-one processor to one-input-to
multiple-output processor, Yu et al proposed an approximate and
probabilistic programmable multi-output quantum processor
\cite{Yu}. Schuch et al proposed a general programmable network to
implement arbitrary two-qubit phase-shift operation with a network
of one-qubit and two- qubit operations \cite{Schuch}. It embodies
the following transformation $|x\ket\rightarrow
e^{-i\theta_x}|x\ket$ where $x$ runs over all possible basis
states. Other interesting results were also obtained
\cite{Vidal,Rosko,Pable,Barenco,Vandersypen}.

In quantum computation, a special controlled phase gate is of
particular importance: it only changes the phase for a particular
basis state $|s\ket$, and leaves the other basis states unchanged.
For instance, in the Grover quantum search
algorithm\cite{Grover97}, there are two phase inversions, $I_\tau$
which inverses the phase of the marked state and leaves alone the
phases of other basis states, and $I_0$ which acts likewise.
Starting from this controlled phase gate, other controlled gate
operation, such as the multi-qubit controlled-not gate can be
realized, and multi-qubit controlled gate is the major operation
in many practical application such as initializing a quantum
computer from $|00\cdots 0\ket$ into an arbitrary superposed
state\cite{longsun}, and in designing quantum cloning
machines\cite{knight}. This multi-qubit gate can be decomposed
into a set of one-qubit gate and two-qubit gates as given in
Ref.\cite{Barenco}. In fact, this decomposition has been used in,
for instance, Refs.\cite{jcplong,Du} to implement multi-qubit
controlled-NOT gate. In the same spirit, the multi-qubit
controlled phase gate can be implemented systematically using the
method proposed in Ref. \cite{Schuch}.

  In this paper, we propose a different realization scheme for the
$n$ -qubit controlled phase gate using modular structure and
recursion. The network retains the same form when the system
extends to more qubits. It contains an $n$-qubit controlled phase
rotation about the $z$-axis plus an $n-1$- qubit  controlled phase
gate. Again the $n-1$- qubit controlled phase gate can then be
further realized similarly. In one hand, the controlled phase gate
is used in many computing algorithms, and it merits special
attention. In the other hand, the modular realization of this
controlled phase gate may provide convenience for some physical
systems such as the nuclear magnetic resonance(NMR). Das et al
proposed the scheme to implement the controlled phase-shift gate
in the NMR system using line-selective pulses, where a controlled
phase rotation was realized through three line selective pulses
\cite{Das}. A line-selective pulse is the high-selective
radio-frequency pulse designed to perturbed transverse
magnetization one line at a time \cite{Linden98}. Experimentally,
line-selective pulses can simplify a network circuit greatly, and
makes it easy to extend the circuit to more qubits. The
operational time of a line-selective pulse is not influenced by
the number of the qubits but the spin system with well resolved
couplings \cite{Du}. For some certain systems, the line-selective
pulses can use short time and this is an attractive feature
considering the limited decoherence time in candidate quantum
computer realization systems. Das et al demonstrated their scheme
in a three qubit NMR system by simulations and in a two qubit
system experimentally. Their method can be applicable for both the
weakly-coupled spin systems and the strongly coupled systems.

  Our work concentrates on the weakly-coupled spin systems, which
have been widely used in quantum computation \cite{Chuang}. We
have experimentally demonstrated the feasibility of the proposed
network in a three qubit NMR quantum system, where the multi-qubit
controlled phase rotation is realized by the line-selective
radio-frequency pulses and the spin selective pulses on the target
qubit. Compared with the method used in Das et al's work, our
method is experimentally simple. In our experiment, the
line-selective pulse that pin-points directly the right frequency
for the controlled configuration has been successfully
demonstrated. The line-selective pulse has the advantage of the
reduced operating time and ease in algorithm design.

The paper is organized as follows. In section \ref{s2}, we
describe the modularized realization of the multi-qubit controlled
phase gate. In section \ref{s3}, we report the details of the
experimental realization in a three-qubit NMR quantum system. In
section \ref{s4}, we discuss some issues in the realization and a
briefly summarize the result.

\section{Modularized quantum network for controlled phase gate}
\label{s2}

   The controlled phase- shift gate is one of most important operation in
quantum computation. On one hand, with a two-bit phase gate and
one-qubit rotation gate, one can build any arbitrary unitary
operation in quantum computation. On the other hand, multi-qubit
controlled phase gate is the essential operation in many quantum
algorithms, such as the Grover quantum search algorithm, quantum
Fourier transform and so
on\cite{Grover97,Grover98,Biham,Weinstein}. The controlled phase-
shift operation for $n$-qubit system is denoted by
$I_{|s_{n}\rangle}^{\varphi }$, where $|s_{n}\rangle$ is a basis
state of an $n$-qubit system. $I_{|s_{n}\rangle}^{\varphi}$
transforms $|s_{n}\rangle$ to $e^{i\varphi }|s_{n}\rangle$, and
leaves the other basis states unaltered. Without loss of
generality, we assume that qubit $n$ is the target bit, and the
other $n-1$ qubits are the controlling bits. Let
$|s_{n}\rangle=|1\cdots 1\rangle$, denoting that all qubits lie in
state $|1\rangle$. $I_{|s_{n}\rangle}^{\varphi }$ can be expressed
as
\begin{equation}\label{02}
 I_{|s_{n}\rangle}^{\varphi }=\left(\begin{array}{cc}
   I_{2^{n}-2} & 0 \\
  0 & U \\
\end{array}\right),
\end{equation}
where
\begin{equation}\label{03}
 U=\left(\begin{array}{cc}
   1 & 0 \\
  0 & e^{i\varphi } \\
\end{array}\right),
\end{equation}
and $I_{2^{n}-2}$  denotes the $(2^{n}-2)\times(2^{n}-2)$ unit
matrix. $U$ is an operation applied to the target qubit. $U$
cannot be constructed from the Pauli operators $\sigma_{x}$,
$\sigma_{y}$, $\sigma_{z}$ and $I_{2}$ directly, because it is not
in $SU(2)$ group\cite{Barenco}. Practically, $U$ can be realized
in the following manner, noting that
\begin{equation}\label{04}
 U=\left(\begin{array}{cc}
   e^{i\varphi /2} & 0 \\
  0 & e^{i\varphi /2}\end{array}\right)
  \left(\begin{array}{cc}
   e^{-i\varphi /2} & 0 \\
  0 & e^{i\varphi /2} \\
\end{array}\right)
\equiv \Phi (\varphi /2)R_{z}(-\varphi ),
\end{equation}
where $R_{z}(-\varphi )\equiv e^{-i\varphi I_{z}}$. $I_{z}$ is
$\hat{z}$-component of the angular momentum of the spin, and
$I_{z}|0\rangle={1\over 2}|0\rangle$, $I_{z}|1\rangle=-{1\over
2}|1\rangle $, by setting $\hbar=1$. $R_{z}(-\varphi)$ can be
realized by a rotation about the $z$-axis, and in NMR it is
realized by radio-frequency (rf) pulses\cite{Linden}. For a single
qubit, the difference between $R_{z}(-\varphi)$ and $U$ is
negligible because they differ only by a global phase. But for the
controlled phase gate, this difference cannot be ignored. For
instance, in a two-qubit system, the controlled-rotation
$R^C_z(\varphi)$ is
\begin{eqnarray}
R^C_z(\varphi)=\left({\begin{array}{cccc}
1 & 0 & 0 & 0\\
0& 1 & 0 & 0\\
0 & 0 & e^{-i\varphi/2} & 0\\
0 & 0 & 0 & e^{i\varphi/2}\end{array}}\right),
\end{eqnarray}
and it is obvious the overall phase between $U$ and $R_z$ is no
longer trivial, and  $\Phi (\varphi /2)$ must be considered in
designing the network for $I_{|s_{n}\rangle}^{\varphi }$.

\begin{figure}
\includegraphics[width=2.5in]{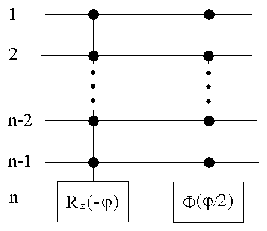}
\caption{The quantum network to implement
$I_{|s_{n}\rangle}^{\varphi}$. The lines represent the $n$ qubits,
respectively. Qubit $n$ is the target qubit, and the other qubits
are the control  ones. $\Phi (\varphi /2)$ is a phase- shift
operation, and $R_{z}(-\varphi )= e^{-i\varphi I_{z}}$.}
\label{fig1}
\end{figure}

\begin{figure}
\includegraphics[width=5in]{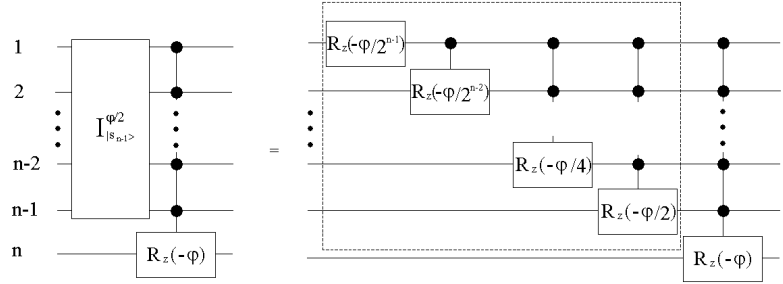}
\caption{The network in Fig. \ref{fig1} which is represented by a
multiple qubit-controlled phase rotation operation and a
compensatory network module, denoted by
$I_{|s_{n-1}\rangle}^{\varphi/2}$, which is equivalent to the
multiple qubit-controlled $\Phi (\varphi /2)$ in Fig. \ref{fig1}.
The compensatory network module can be further decomposed as the
network circuit denoted by the dashed rectangle in the right
network. }\label{fig2}
\end{figure}

The network to implement $I_{|s_{n}\rangle}^{\varphi }$ is shown
by Fig. \ref{fig1}, where the "global" controlled-$\Phi(\varphi
/2)$ is explicitly drawn.  By a simple observation, it is not
difficult to show that the controlled-$\Phi(\varphi/2)$ is
equivalent to:
\begin{eqnarray}
 \Phi^{C\cdots C}(\varphi/2)=I_{|s_{n-1}\rangle}^{\varphi/2},
 \end{eqnarray}
where $I_{|s_{n-1}\rangle}^{\varphi/2}$ denotes the phase- shift
operation for the $n-1$- qubit system composed of the control
qubits in the original $n$-qubit system. $|s_{n-1}\ket$ is the
basis state of the first $n-1$ control qubits where the
corresponding basis state of
 the whole $n$ qubit is $|s_{n}\rangle=|s_{n-1}\rangle|1\rangle_{n}$.
 The $n-1$-th qubit is the
target qubit, the other $n-2$ qubits are the control qubits in
$I_{|s_{n-1}\rangle}^{\varphi/2}$. The effect of this shortened
controlled phase gate can be seen as follows: the controlled-phase
rotation $R^C_z(\varphi)$ changes
\begin{eqnarray}
|s_{n-1}0\ket&& \rightarrow  e^{-i\varphi/2}
|s_{n-1}0\ket,\nonumber\\
 |s_{n-1}1\ket && \rightarrow
e^{i\varphi/2} |s_{n-1}1\ket,
\end{eqnarray}
and it leaves other basis states unaltered. A follow-up
$I_{|s_{n-1}\ket}^{\varphi/2}$ operation changes $|s_{n-1}\ket$
into $e^{i\varphi/2}|s_{n-1}\ket$, and this takes $e^{-i\varphi/2}
|s_{n-1}0\ket$ into $ |s_{n-1}0\ket$, and $e^{i\varphi/2}
|s_{n-1}1\ket$ into $e^{i\varphi} |s_{n-1}1\ket$. Hence the total
effect is to change $|s_{n}\ket$ into $e^{i\varphi}|s_{n}\ket$.
Fig. \ref{fig1} can be further represented as Fig. \ref{fig2},
where $I_{|s_{n-1}\rangle}^{\varphi/2}$ is applied to the control
qubits. We call the $I_{|s_{n-1}\rangle}^{\varphi/2}$ part as the
compensatory network module. The controlled-phase rotation can be
realized by two multi-qubit controlled-NOT gates ($\Lambda
_{n-1}({\rm NOT})$) and two single-bit phase rotations
\cite{Barenco}.  $\Lambda _{n-1}({\rm NOT})$ can be directly
realized in some physical realizations such as the NMR system, or
be further reduced to a set of  single qubit rotations and
two-qubit controlled NOT gates as described in Ref.\cite{Barenco}.

The nice feature in Fig.\ref{fig2} is that
$I_{|s_{n-1}\ket}^{\varphi/2}$ can be realized in the same way as
$I_{|s_n\ket}^{\varphi}$: it can be realized by a compensatory
module composed of the $n-2$ qubits together with an $n-2$ qubits
controlled phase rotation, and the compensatory module unit can be
similarly constructed shown as the right network in Fig.
\ref{fig2}. Hence we have obtained a recursive way to construct
the multi-qubit phase shift gate.  In summary,
$I_{|s_{n}\rangle}^{\varphi}$ is constructed by the n-1 -qubit
controlled phase rotation and the compensatory network module
$I_{|s_{n-1}\rangle}^{\varphi/2}$. $I_{|s_{n}\rangle}^{\varphi}$
is realized recursively.

\section{NMR realization of the modularized network}
\label{s3}

   In the experiment, we use a sample of Carbon-13
labelled trichloroethylene (TCE) dissolved in d-chloroform. Data
are taken at controlled temperature (22$^{0}$) with a Bruker DRX
500 MHz spectrometer. $^{1}$H is denoted as qubit 3, the $^{13}$C
directly connecting to $^{1}$H is denoted as qubit 2, and the
other $^{13}$C is denoted as qubit 1. The three qubits are denoted
as C1, C2 and H3. By setting $\hbar=1$, the Hamitonian of the
three-qubit system is
\begin{equation}\label{hamidun}
  H=-2\pi\nu_{1}I_{z}^{1}-2\pi\nu_{2}I_{z}^{2}-2\pi\nu_{3}I_{z}^{3}
  +2\pi J_{12}I_{z}^{1}I_{z}^{2}+2\pi J_{23 }I_{z}^{2}I_{z}^{3}
  +2\pi J_{13} I_{z}^{1}I_{z}^{3},
\end{equation}
where $\nu_{1}$, $\nu_{2}$, $\nu_{3}$ are the resonance
frequencies for C1, C2 and H3, respectively, and
$\nu_{1}=\nu_{2}+904.4$Hz. The coupling constants are measured to
be $J_{12}=103.1$ Hz, $J_{23}=203.8$ Hz, and $J_{13}=9.16$ Hz
respectively.

   We realize $I_{|11\rangle}^{\varphi}$ in the two-qubit system
composed of C1 and C2, through decoupling H3. The spin-coupling
evolution between C1 and C2 is described by
\begin{equation}\label{ouhe}
  [\tau]=e^{-i2\pi J\tau I_{z}^{1} I_{z}^{2}}.
\end{equation}
C1 and C2 construct a homonuclear system, which can demonstrate
the implementation of $I_{|11\rangle}^{\varphi}$ clearly, and
represent the effect of the compensatory module. Fig. \ref{fig2}
is simplified to a two-qubit network, where C1 is the control
qubit, and C2 is the target qubit. The compensatory module is
expressed as $I_{|1\rangle}^{\varphi/2}=\Phi^{1}(\varphi
/4)R_{z}^{1}(-\varphi /2)$, and is equivalent to
$R_{z}^{1}(-\varphi /2)$. $\Phi^{1}(\varphi/4)$ only contributes
an irrelevant overall phase factor before
$I_{|11\rangle}^{\varphi}$. $\Lambda _{n-1}(NOT)$ is the ordinary
CNOT operation. The operation sequence CNOT-$R_{z}^{2}(\varphi
/2)$-CNOT can be realized by $[\frac{-\varphi}{2\pi J_{12}}]$.
Hence $I_{|11\rangle}^{\varphi}$ is realized by the following
pulse sequence
$$R_{z}^{1,2}(-\varphi /2)-[\frac{-\varphi}{2\pi J_{12}}],$$
where $\varphi$ is taken as the menus value during the experiment.

   The experiment starts with the pseudo-pure state prepared by spatial averaging.
The following pulse sequence \cite{Somarooprl991,Fang}
$$[\frac{\pi}{4}]_{x}^{1,2}-\frac{1}{4J_{12}}-[\pi]_{y}^{1,2}
-\frac{1}{4J_{12}}-[-\pi]_{y}^{1,2}
-[-\frac{5\pi}{6}]_{y}^{1,2}-[grad]_{z}$$ transforms the system
from equilibrium state to the pseudo-pure state
\begin{equation}\label{pure}
    \rho_{ini}=I_{z}^{1}/2+I_{z}^{2}/2+I_{z}^{1}I_{z}^{2},
\end{equation}
which is equivalent to state $|00\rangle$. The symbol $1/4J_{12}$
denotes the evolution caused by the magnetic field for $1/4J_{12}$
when the pulses are switched off. After the application of the
Hadamard transform, $|00\rangle$ is transform to the superposition
of states
\begin{equation}\label{ss}
   |p\rangle=(|00\rangle+|01\rangle+|10\rangle+|11\rangle)/2.
\end{equation}
It is obvious $I_{|11\rangle}^{\varphi}|p\rangle=
(|00\rangle+|01\rangle+|10\rangle+e^{i\varphi}|11\rangle)/2$,
represented by density matrix
\begin{equation}\label{state2}
 \rho_{f}=\frac{1}{4}\left( \begin{array}{cccc}
   1&1 & 1 & e^{-i\varphi} \\
   1 & 1 & 1 & e^{-i\varphi} \\
   1 & 1 & 1 & e^{-i\varphi} \\
   e^{i\varphi} & e^{i\varphi} & e^{i\varphi} & 1 \\
 \end{array}\right).
\end{equation}
In Eq. (\ref{state2}), the matrix elements $\rho_{f}(1,3)$ and
$\rho_{f}(2,4)$ can be directly observed through the spectra of
C1,  and $\rho_{f}(1,2)$, $\rho_{f}(3,4)$  can be directly
observed through the spectra of C2.

\begin{figure}
\includegraphics[width=5in]{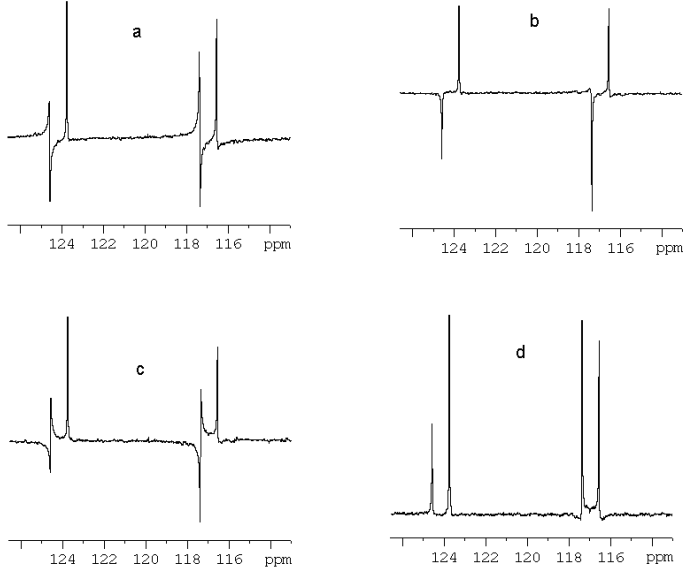}
\caption{$^{13}C$ spectra of trichloroethylene after
$I_{|s_{n}\rangle}^{\varphi}$ is applied to the uniform
superposition of basis states obtained through applying Hadmamard
transform to $|00\rangle$. $^{1}H$ has been decoupled. The phases
of the right peaks of in the doublets hardly change. The phases of
the left peaks, denoted by $Ph2$ (2 for the two qubit system),
change with $-\varphi/2$ proportionably. Figs. (a-d) are the
spectra for $-\varphi/2=\pi/4$, $\pi/2$, $3\pi/4$ and $\pi$,
respectively.} \label{c1c2}
\end{figure}

\begin{figure}
\includegraphics[width=4in]{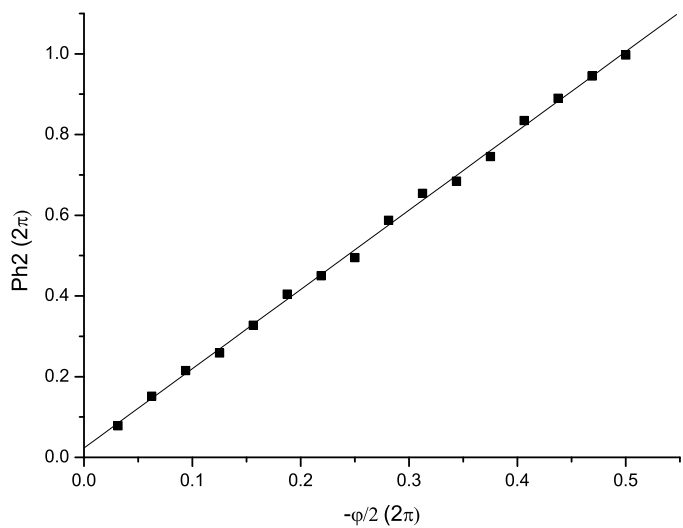}
\caption{The phase of the left peak in the doublet of C1 or C2,
which is denoted as $Ph2$, versus $-\varphi/2$. The graph can be
fitted as a line with slope equal to 1.96.  The theoretical
expectation is 2.} \label{phase1}
\end{figure}

   The Hadamard transform is realized by $[-\frac{\pi}{2}]_{y}^{1,2}-[\pi]_{x}^{1,2}$.
$I_{|11\rangle}^{\varphi}$ is realized by
$$\frac{-\varphi}{4\pi J_{12}}-[\pi]_{y}^{1,2}-\frac{-\varphi}{4\pi J_{12}}-
[-\pi]_{y}^{1,2}-[-\frac{\pi}{2}]_{y}^{1,2}-[-\frac{\varphi}{2}]_{x}^{1,2}-
[\frac{\pi}{2}]_{y}^{1,2}.$$ Figs. \ref{c1c2}(a-d) show the
results when $-\varphi/2=\pi/4$, $\pi/2$, $3\pi/4$ and $\pi$,
respectively. In each figure, the frequency centers of the
doublets of C1 and C2 are 124.14ppm and 116.91ppm, respectively.
The right and left peaks in the doublet of C1 correspond to
$\rho_{f}(1,3)$ and $\rho_{f}(2,4)$ in Eq.(\ref{state2}), and the
right and left ones in the doublet of of C2 correspond to
$\rho_{f}(1,2)$ and $\rho_{f}(3,4)$ in Eq.(\ref{state2})
respectively. The phases of the right peaks of C1 and C2 hardly
change with $-\varphi/2$. They are chosen as the reference phases
to observe the phases of the other peaks so that the phases of
signals are meaningful\cite{Jones}. The phase of the left peak
(denoted by $Ph2$) changes via $-\varphi/2$ proportionably, which
are shown in Fig. \ref{phase1}.

   To show the effect of the compensatory module, we add $R^{1}_{z}(\varphi/2)$
   at the end of network shown in Fig.
\ref{fig2} to cancel the compensatory network module. The network
without the compensatory network module implements the
transformation
\begin{equation}\label{Isp}
  I_{|11\rangle}^{\varphi}R^{1}_{z}(\varphi/2)=e^{i\varphi/4}\left( \begin{array}{cccc}
   1&0 & 0 & 0 \\
   0 & 1 & 0 & 0 \\
   0& 0 & e^{-i\varphi/2} & 0 \\
   0 & 0 & 0 & e^{i\varphi/2} \\
 \end{array}\right).
\end{equation}
It is obvious that
$I_{|11\rangle}^{\varphi}R^{1}_{z}(\varphi/2)|p\rangle
=e^{i\varphi/4}(|00\rangle+|01\rangle+e^{-i\varphi/2}|10\rangle+e^{i\varphi/2}|11\rangle)/2$,
which is represented as
\begin{equation}\label{statep}
 \rho_{f}^{'}=\frac{1}{4}\left( \begin{array}{cccc}
   1 &1 & e^{i\varphi/2} & e^{-i\varphi/2} \\
   1 & 1 & e^{i\varphi/2} & e^{-i\varphi/2} \\
   e^{-i\varphi/2} & e^{-i\varphi/2} & 1 & e^{-i\varphi} \\
   e^{i\varphi/2} & e^{i\varphi/2} & e^{i\varphi} & 1 \\
\end{array}\right).
\end{equation}
From Eq. (\ref{statep}), the signals of C1 and C2 have phase
difference, which can be observed in this homonuclear system.

\begin{figure}
\includegraphics[width=6in]{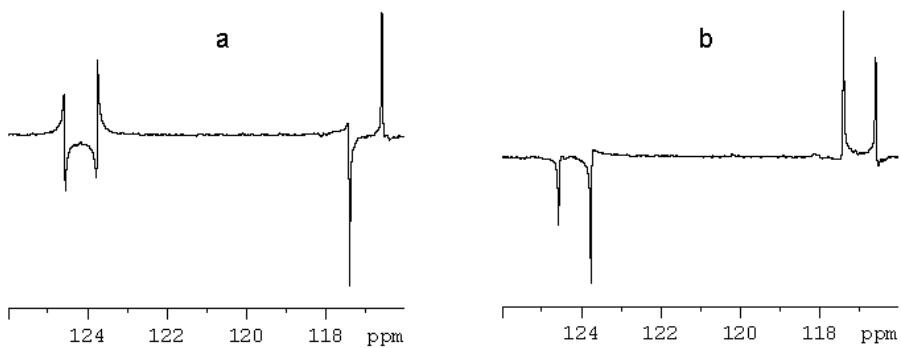}
\caption{ $^{13}C$ spectra after
$I_{|11\rangle}^{\varphi}R^{1}_{z}(\varphi/2)$ is applying to the
uniform superposition of basis states, when $-\varphi/2=\pi/2$,
$\pi$, shown by Figs. (a) and (b), respectively.} \label{143}
\end{figure}

   $R^{1}_{z}(\varphi/2)$ can be realized by
$$ t -[\pi]_{y}^{2}-2t-[-\pi]_{y}^{2}- t.$$
In theory, $\varphi/2=4(\nu_{1}-\nu_{2})2\pi(t+t_{p}/2)$, where
$t_{p}$ is the  width of the $\pi$  pulse selective for C2
\cite{Linden}. However, the dependence of $\varphi/2$ on $t$ needs
be measured due to the errors caused by decoherence and
imperfection in the pulses. Figs. \ref{143}(a-b) show the results
of $I_{|11\rangle}^{\varphi}R^{1}_{z}(\varphi/2)|p\rangle$ with
$-\varphi/2=\pi/2$, $\pi$. Compared with Figs. \ref{phase1}(b) and
(d), consequently the phases in peaks C1 and C2 are no longer the
same. Theoretically this phase difference should be $\varphi/2$,
and our experiment results agree with this theoretical
expectation.

\section{Discussion and summary}
\label{s4}

\begin{figure}
\includegraphics[width=2.5in]{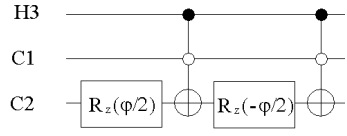}
\caption{The quantum network to implement
$I^{-\varphi/2}_{|001\rangle} I^{\varphi/2}_{|011\rangle}$.}
\label{network3}
\end{figure}

  We realize the network shown in Fig. \ref{network3} in a three qubit
system by switching off the decoupling pulses for $H3$.  The
network consists of two rotation operations for C2 and two
controlled-controlled-NOT gates (CCNOT or Toffoli gate). C2 is the
target qubit, C1 and H3 are the control qubits  with control
condition $|0\rangle_{1}|1\rangle_{3}$. The network implements the
transform $I^{-\varphi/2}_{|001\rangle}
I^{\varphi/2}_{|011\rangle}$, which can be transformed to
$I^{\varphi}_{|011\rangle}$ by multiplying
$I^{\varphi/2}_{|001\rangle} I^{\varphi/2}_{|011\rangle}$. In
fact, $I^{\varphi/2}_{|001\rangle} I^{\varphi/2}_{|011\rangle}$ is
the controlled phase- shift operation for the system composed of
C1 and H3, and it is represented as

\begin{equation}\label{Is2}
I_{|0\rangle_{1}|1\rangle_{3}}^{\varphi/2}=\left(\begin{array}{cccc}
    1 & 0 & 0 & 0 \\
    0& e^{i\varphi/2} & 0 & 0 \\
    0 & 0 & 1 & 0 \\
    0 & 0 & 0 & 1 \
  \end{array}\right),
\end{equation}
with the basis state order $|0\rangle_{1}|0\rangle_{3}$,
$|0\rangle_{1}|1\rangle_{3}$, $|1\rangle_{1}|0\rangle_{3}$,
$|1\rangle_{1}|1\rangle_{3}$.
$I_{|0\rangle_{1}|1\rangle_{3}}^{\varphi/2}$ can realized by a
similar way in Sec. \ref{s3}. $I^{-\varphi/2}_{|001\rangle}
I^{\varphi/2}_{|011\rangle}$ transforms the superposition of
states

\begin{equation}\label{p3}
 |p\rangle_{3}=(|000\rangle-|001\rangle -|010\rangle+|011\rangle)/2
\end{equation}
to $(|000\rangle-e^{-i\varphi/2}|001\rangle
-|010\rangle+e^{i\varphi/2}|011\rangle)/2$, expressed as the
density matrix

\begin{equation}\label{pf3}
    \rho_{f3}=\left(
    \begin{array}{cccccccc}
      1 & -e^{i\varphi/2} & -1 & e^{-i\varphi/2} & 0 & 0 & 0 & 0 \\
      -e^{-i\varphi/2} & 1 & e^{-i\varphi/2} & -e^{-i\varphi} & 0 & 0 & 0 & 0 \\
      -1 & e^{i\varphi/2} & 1 & -e^{-i\varphi/2} & 0 & 0 & 0 & 0 \\
      e^{i\varphi/2}  & -e^{i\varphi} &  -e^{i\varphi/2} &1 & 0 & 0 & 0 & 0 \\
      0 & 0 & 0 & 0 & 0 & 0 & 0 & 0 \\
      0 & 0 & 0& 0 & 0 & 0 & 0 & 0 \\
      0 & 0 & 0 & 0 & 0 & 0 & 0 & 0 \\
      0 & 0 & 0 & 0 & 0 & 0 & 0 & 0 \\
    \end{array}
  \right),
\end{equation}
where $|p\rangle_{3}$ is obtained by applying
$[\frac{\pi}{2}]_{y}^{2,3}$ to $|000\rangle$. The elements (1,3)
and (2,4) in the matrix can be observed in the spectrum of C2
directly.

  The pseudo-pure state $|000\rangle$ can be represented by

\begin{equation}\label{eff}
    \rho_{eff}=(\frac{1}{2}I+I_{z}^{1})(\frac{1}{2}I+I_{z}^{2})(\frac{1}{2}I+I_{z}^{3}).
\end{equation}
$\rho_{eff}$ can be expressed as $\rho_{eff} \equiv
\rho_{1}+\rho_{2}+\frac{1}{8}I_{8\times8}$, where

  \begin{equation}\label{eff1}
\rho_{1}
=(\frac{1}{2}I_{z}^{1}+\frac{1}{2}I_{z}^{2}+I_{z}^{1}I_{z}^{2}+\frac{1}{4})I_{z}^{3},
\end{equation}

\begin{equation}\label{eff2}
\rho_{2}
=\frac{1}{2}(\frac{1}{2}I_{z}^{1}+\frac{1}{2}I_{z}^{2}+I_{z}^{1}I_{z}^{2}),
\end{equation}
and the last term $\frac{1}{8}I_{8\times8}$ can be ignored
\cite{Knill1}. Quantum computation can start with $\rho_{1}$ and
$\rho_{2}$, respectively. The average of the two results is
equivalent to the result obtained from the initial state
$\rho_{eff}$. Because the results of applying
$[\frac{\pi}{2}]_{y}^{2,3}$ to $\rho_{1}$ have no observable
signals in the carbon spectra, it is sufficient to use $\rho_{2}$
 as the initial state to observe the results of applying
$I^{-\varphi/2}_{|001\rangle} I^{\varphi/2}_{|011\rangle}$ to
$|p\rangle_{3}$. This fact can simplify the process of experiments
greatly.

\begin{figure}
\includegraphics[width=6in]{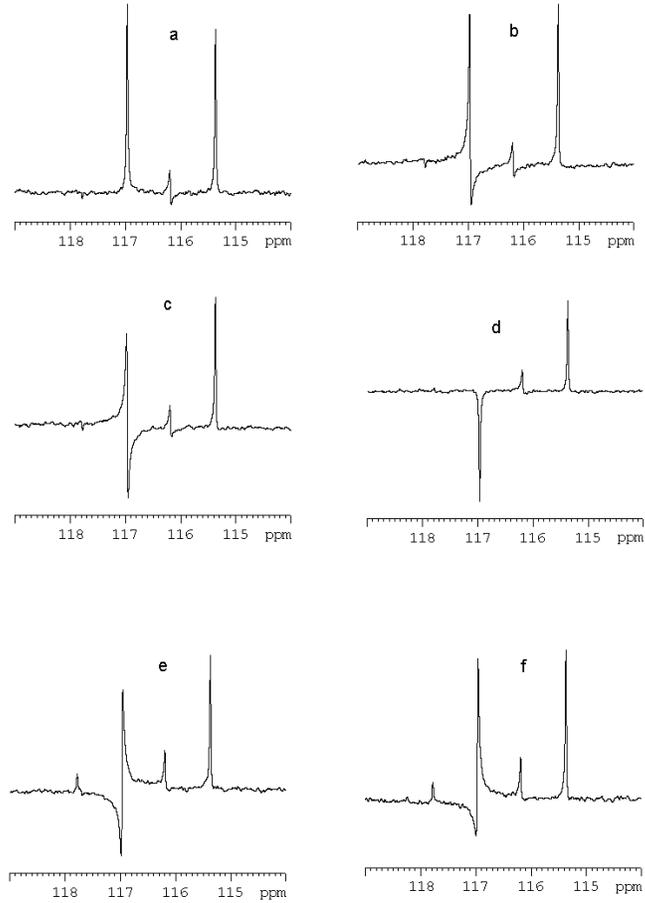}
\caption{Spectra of C2 after $I^{-\varphi/2}_{|001\rangle}
I^{\varphi/2}_{|011\rangle}$ is applied to
$(|000\rangle-|001\rangle -|010\rangle+|011\rangle)/2$. Only two
NMR peaks appear in each spectrum. The phase of the right peak
hardly changes. The phase of the left peak changes with
$\varphi/2$, proportionably. Figs. (a-f) show the spectra for
$\varphi/2=0$, $25.2^{\circ}$,  $38.2^{\circ}$, $90^{\circ}$,
$141.8^{\circ}$,  $154.7^{\circ}$,  respectively.} \label{C2nmr}
\end{figure}

\begin{figure}
\includegraphics[width=4in]{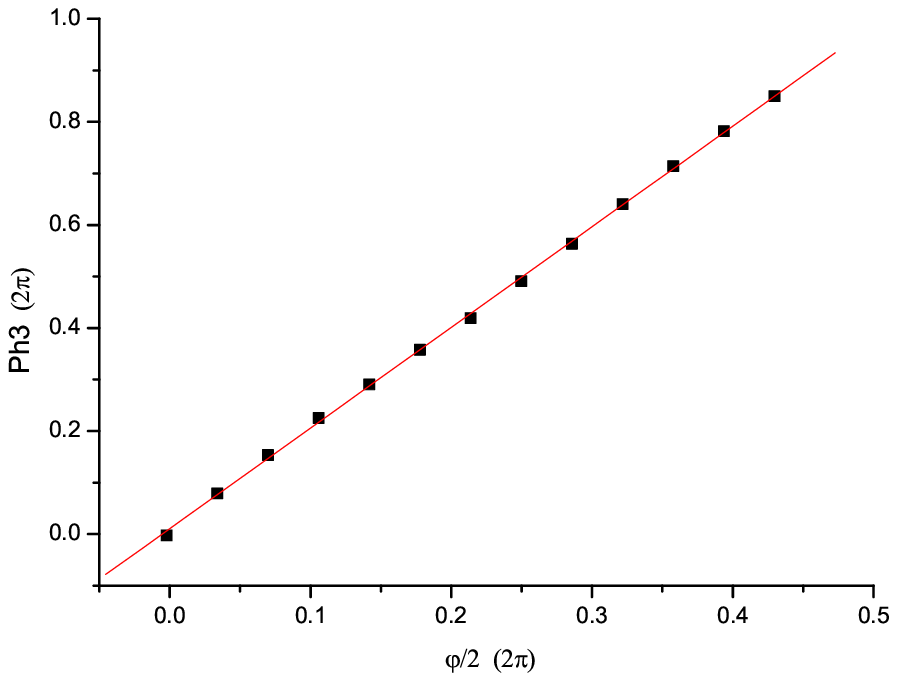}
\caption{The phase of the left peak in the spectrum shown in Fig.
\ref{C2nmr}, which is denoted as $Ph3$, versus $\varphi/2$. The
graph can be fitted as a line with slope approximate to 1.95. In
theory, slope is -2. The difference of sign results from the
approximation of CCNOT gates.} \label{theta-phi}
\end{figure}

 The initial state described as Eq.
(\ref{eff2}) is realized by
$[\frac{\pi}{2}]_{y}^{3}-[grad]_{z}-[\frac{\pi}{4}]_{x}^{1,2}-\frac{1}{8J_{12}}-
[\pi]_{y}^{3}-\frac{1}{8J_{12}}-[\pi]_{y}^{1,2,3}-\frac{1}{8J_{12}}-
[-\pi]_{y}^{3}-\frac{1}{8J_{12}}-[-\pi]_{y}^{1,2,3}-[-\frac{5\pi}{6}]_{y}^{1,2}-[grad]_{z}$.
CCNOT gate can be approximately realized by a line selective $\pi$
pulse with frequency $\nu_{2}+J_{12}/2-J_{23}/2$, which causes
transition $|001\rangle\leftrightarrow |011\rangle$. Figs.
\ref{C2nmr} show the results of
$I^{-\varphi/2}_{|001\rangle}I^{\varphi/2}_{|011\rangle}$$|p\rangle_{3}$.
In each figure, the right peak, with frequency
$\nu_{2}+J_{12}/2+J_{23}/2$, corresponds the element (1,3), and
the left peak, with frequency $\nu_{2}+J_{12}/2-J_{23}/2$,
corresponds the element (2,4), which changes versus $\varphi/2$.
Figs. \ref{C2nmr} (a-f) show the results of $\varphi/2=0$,
$25.2^{\circ}$,  $38.2^{\circ}$, $90^{\circ}$,  $141.8^{\circ}$,
$154.7^{\circ}$,  respectively. The graph in Fig. \ref{theta-phi}
shows the phase of the right peak (denoted by $Ph3$) versus
$\varphi/2$. It can be fitted as a line with slope approximate to
1.95. In theory, slope is -2. The difference of sign results from
the approximation of CCNOT gates.

   The modularized quantum network to implement the
controlled phase- shift gate is proposed, and realized on a three
qubit NMR quantum computer. The $n$- qubit network nests the
$n-1$- qubit network. This idea represents the essential thoughts
of computer science. The quantum modules provide a systemic method
to building practical quantum computers. It is sure that the
modularized quantum networks will play central role for large-
scale quantum computer.

\section* {Acknowledgements}
   This work is supported by the National Natural Science
Foundation of China under Grant No. 10374010, 60073009, 10325521,
the National Fundamental Research Program Grant No. 001CB309308,
the Hang-Tian Science Fund, the SRFDP program of Education
Ministry of China, and  China Postdoctoral Science Foundation.

\newpage


\begin{thebibliography}{}

\bibitem{shor} P W Shor,  Proc. 35th Annual IEEE Symposium on Foundations of
  Computer Science-FOCS, 20-22 (1994)

\bibitem{Grover97}L. K. Grover, Phys. Rev. Lett. 79, 325 (1997)


\bibitem{Deutsch85}D. Deutsch, Proc. R. Soc. A, 400, 97-117(1985)

\bibitem{Deutsch95} D. Deutsch, A. Barenco, and A. Ekert, Proc. R.
    Soc. London A 449, 669 (1995); quant-ph/9505018

\bibitem{Bremner} M. J. Bremner, C. M. Dawson, J. L. Dodd,  A. Gilchrist,
          A. W. Harrow, D. Mortimer, M. A. Nielsen, and T. J.
          Osborne1, Phys. Rev. Lett, 89, 247902 (2002)

\bibitem{Nielsen} M. A. Nielsen, and I. L. Chuang, {\sl Quantum computation and
 quantum information(Cambridge University Press,2000)}

\bibitem{Barenco} A. Barenco, C. H. Bennett, R. Cleve, D. P. DiVincenzo, N. Margolus,
   P. Shor, T. Sleator, J. A. Smolin, and H. Weinfurter, Phys. Rev. A, 52, 3457(1995)

\bibitem{Cleve} R. Cleve, A. Ekert, C. Macchiavello, and M. Mosca,
        Proc. R. Soc, Lond, A 454, 339 (1998)

\bibitem{Nielsen97} M. A. Nielsen, and I. L. Chuang, Phys. Rev. Lett. 79, 321 (1997)

\bibitem{Yu} Y.-F. Yu, J. Feng, and M. -S. Zhan, Phys. Rev. A 66, 052310 (2002)

\bibitem{Schuch} N. Schunch, and J. Siewert,
                 Phys. Rev. Lett, 91, 027902(2003)

\bibitem{Vidal} G. Vidal, L. Masanes, and J. I. Cirac, Phys. Rev. Lett. 88, 047905 (2002)

\bibitem{Rosko} M. Ro$\check{s}$ko, V. Bu$\check{z}$ek, P. R.
        Chouha, and M. Hillery, Phys. Rev. A 68, 062302 (2003)

\bibitem{Pable} J. P. Paz, and A. Roncaglia, Phys. Rev. A 68, 052316 (2003)

\bibitem{Vandersypen} L. M. K. Vandersypen, M. Steffen, M. H. Sherwood, C. S. Yannoni,
    G. Breyta, and I. L. Chuang, Appl. Phys. Lett, 76, 646(2000)



\bibitem{longsun} G. L. Long and Y. Sun, Phys. Rev. A 64, 014303
(2001)

\bibitem{knight} K. Maruyama and P. L. Knight, Phys. Rev. A 67,
032303 (2003)

\bibitem{jcplong} G. L. Long and L. Xiao, J. Chem. Phys. 119, 8473
(2003)

\bibitem{Das} R. Das, T. S. Mahesh, and A. Kumar, J. Magn. Res.
159, 46 (2002)

\bibitem{Linden98} N. Linden, H. Barjat, R. Freeman,  Chem. Phys.
    Lett, 296, 61(1998)

\bibitem{Du} J.-F. Du, M.-J. Shi, J. -H. Wu, X.-Y. Zhou, and R.-D. Han,
 Phys. Rev. A, 63, 042302(2001)

\bibitem{Chuang} L. M. K. Vandersypen, and I. L. Chuang,
quant-ph/0404064

\bibitem{Grover98} L. K. Grover, Phys. Rev. Lett, 80, 4329(1998)

\bibitem{Biham} E. Biham, O. Biham ,D. Biron , M. Grassl , D. A. Lidar, and
            D. Shapira, Phys. Rev. A, 63, 012310 (2000)

\bibitem{Weinstein} Y. S. Weinstein, M. A. Pravia, E. M. Fortunato, S. Lloyd,  and D. G. Cory,
   Phys. Rev. Lett, 86, 1889(2001)

\bibitem{Linden} N. Linden, B. Herv$\grave{e}$, R. J. Carbajo, and R. Freeman, Chem. Phys. Lett,
    305, 28(1999)


\bibitem{Somarooprl991} S. Somaroo, C. H. Tseng, T. F. Havel, R. Laflamme,
       and D. G. Cory, Phys.  Rev. Lett, 82, 5381(1999)

\bibitem{Fang}F.-X. Mang, X.- W. Zhu , M. Feng  , X.- A. Mao, and F. Du,
             Phys. Rev. A, 61, 022307(2000)

\bibitem{Jones} J.  A. Jones, in {\sl The Physics of
              quantum Information}, edited by D. Bouwmeester, A. Ekert, and A. Zeilinger
              (Springer, Berlin Heidelberg, 2000) pp.177-189.




\bibitem{Knill1} E. Knill, I. Chuang, and R. Laflamme,  Phys. Rev. A, 57, 3348(1998)


\end{thebibliography}
\end{document}